\documentstyle[epsfig,cite]{mn2e}
\begin{document}

\title[Spectral evolution of GRO J1655$-$40, as observed by {\it Swift}]
      {The 2005 outburst of GRO J1655$-$40: spectral evolution of the rise, as observed by {\it Swift}}
\author[Brocksopp et al.]
    {C.~Brocksopp$^1$\thanks{email: cb4@mssl.ucl.ac.uk}, K.E.~McGowan$^1$, H.~Krimm$^{2,3}$, O.~Godet$^4$, P.~Roming$^5$, K.~O.~Mason$^1$,
\newauthor
 N.~Gehrels$^2$, M.~Still$^{2,3}$, K.~Page$^4$, A.~Moretti$^6$, C.R.~Shrader$^2$, S.~Campana$^{6}$, J.~Kennea$^{5}$\\
$^1$Mullard Space Science Laboratory, University College London, Holmbury St. Mary, Dorking, Surrey RH5 6NT, UK\\
$^2$NASA/Goddard Space Flight Center, Greenbelt, MD 2077, USA\\
$^3$Universities Space Research Association, Columbia, MD 20144 USA\\
$^4$Department of Physics \& Astronomy, University of Leicester, Leicester LE1 7RH, UK\\
$^5$Department of Astronomy and Astrophysics, Pennsylvania State University, 525 Davey Laboratory, University Park, PA 16802, USA\\
$^6$INAF-Osservatorio Astronomico di Brera, Via Bianchi 46, I-23807 Merate (LC), Italy\\
}
\date{Accepted ??. Received ??}
\pagerange{\pageref{firstpage}--\pageref{lastpage}}
\pubyear{??}
\maketitle

\begin{abstract}
We present {\sl Swift} observations of the black hole X-ray transient, GRO J1655$-$40, during the recent outburst. With its multiwavelength capabilities and flexible scheduling, {\sl Swift} is extremely well-suited for monitoring the spectral evolution of such an event. GRO J1655$-$40 was observed on 20 occasions and data were obtained by all instruments for the majority of epochs. X-ray spectroscopy revealed spectral shapes consistent with the ``canonical'' low/hard, high/soft and very high states at various epochs. The soft X-ray source (0.3--10 keV) rose from quiescence and entered the low/hard state, when an iron emission line was detected. The soft X-ray source then softened and decayed, before beginning a slow rebrightening and then spending $\sim 3$ weeks in the very high state. The hard X-rays (14--150 keV) behaved similarly but their peaks preceded those of the soft X-rays by up to a few days; in addition, the average hard X-ray flux remained approximately constant during the slow soft X-ray rebrightening, increasing suddenly as the source entered the very high state. These observations indicate (and confirm previous suggestions) that the low/hard state is key to improving our understanding of the outburst trigger and mechanism. The optical/ultraviolet lightcurve behaved very differently from that of the X-rays; this might suggest that the soft X-ray lightcurve is actually a composite of the two known spectral components, one gradually increasing with the optical/ultraviolet emission (accretion disc) and the other following the behaviour of the hard X-rays (jet and/or corona).
\end{abstract}

\section{Introduction}

{\sl Swift} (Gehrels et al. 2004) was launched on 2004 November 20 and, with its large field of view and rapid triggering ability, is a satellite dedicated to the detection and follow-up of gamma-ray bursts. The Burst Alert Telescope (BAT; Barthelmy et al. 2005) operates at 14--150 keV, has a 2-steradian field of view and provides the trigger information. Once a trigger is generated by the BAT, the spacecraft then slews within a few tens of seconds to point the X-ray Telescope (XRT; Burrows et al. 2005) and the Ultraviolet/Optical Telescope (UVOT; Roming et al. 2005) towards the region of interest, thereby providing broad-band spectral coverage of any detected gamma-ray burst. The XRT operates at 0.3--10.0 keV. The UVOT obtains photometry through three optical and three ultraviolet filters, as well as optical and ultraviolet grism spectra.

The multi-wavelength aspect to {\it Swift} is also extremely valuable for the study of other objects, including black hole X-ray transients which are known to exhibit very different behaviour in each of the various frequency regimes. We would typically expect to detect optical light from the companion star and the accretion disc, and ultraviolet from the irradiated face of the companion star and the accretion disc. Emission in the infra-red is expected from the star and the jet. We further detect radio emission from the jet and X-rays from the accretion disc, Comptonising corona and/or jet (see e.g. McClintock \& Remillard 2005; Fender 2005). Therefore multiwavelength studies are not just preferable, but are actually vital to our understanding of the outbursts which take place.

Black hole X-ray transient outbursts are known to pass through a number of X-ray spectral states over the course of their outbursts; these spectral states are also associated with specific jet behaviour (van der Klis 2005; McClintock \& Remillard 2005; Fender 2005; Homan \& Belloni 2005). Brocksopp et al. (2002a, 2004) emphasized the importance of the initial low/hard state and the necessity of observing the onset of the outburst if we are ever to understand the mechanism by which these outbursts take place. The rapid trigger and Target of Opportunity capabilities of {\sl Swift} make it ideal for detection and/or follow-up of any transient event, particularly those taking place at hard X-ray energies. Therefore, since a period of low/hard state behaviour has been observed at the onset of every black hole transient outburst for which data were available (Brocksopp et al. 2002a), the BAT is much better suited to early observation of these events than the only other currently available All Sky Monitor on-board the Rossi X-ray Timing Explorer {\sl RXTE}.

GRO~J1655$-$40 is one such X-ray transient which was first discovered in 1994 July when it entered an outburst and was detected at 20--100 keV by the Burst and Transient Source Experiment (BATSE) on-board the Compton Gamma Ray Observatory (CGRO) (Harmon et al. 1995).  The sequence of hard X-ray outbursts that followed was well-correlated with the radio observations obtained at the Very Large Array (VLA) (Harmon et al. 1995) but this first outburst of GRO~J1655$-$40 was particularly notable for the discovery of radio jets travelling with apparent superluminal motion (Tingay et al. 1995). This was later confirmed by Hjellming \& Rupen (1995) who studied the episodic ejection nature of the jets; later re-analysis of radio data from the 1994 August-September outbursts revealed 10\% linear polarization, confirming the synchrotron nature of the jets (Hannikainen et al 2000). While GRS~1915+105 had been the first Galactic object to show this phenomenon (Mirabel \& Rodr\'{\i}guez 1994; Fender \& Belloni 2004), its high extinction made optical study very difficult; thus a second source with jet velocity $\ge 0.9c$ was very welcome.

Optical study of GRO~J1655$-$40 was much more straightforward than of GRS~1915+105 and a counterpart was detected at $V\sim 14.4$ magnitudes by Bailyn et al. (1995). Study of older photographic plates showed that there had been a rise of $\sim3$ magnitudes above quiescence. The optical counterpart had fluxes and orbital parameters comparable with those of other black hole candidates; Orosz \& Bailyn (1997) later obtained an orbital period of 2.6 days and a mass for the compact object of 7.0 $M_{\odot}$, confirming its black hole nature, and a mass for the companion star of 2.3 $M_{\odot}$ (but see also Beer \& Podsiadlowski 2002).

Zhang et al. (1997) presented a broad-band X-ray spectrum for the 1995 July-August outburst, compiling simultaneous data from {\sl ASCA}, {\sl Granat}/WATCH and {\sl CGRO}/BATSE and thereby spanning 1 keV -- 2 MeV. The spectrum was described using a multi-colour disc model and higher energy power-law. Additional {\sl ASCA} spectra revealed the presence of absorption features at 6.63 and 6.95 keV, consistent with highly ionized iron lines (Ueda et al. 1998).

The sequence of episodic outbursts continued into 1995, totalling at least five separate peaks (although see also Hannikainen et al. 2000), but interestingly the later X-ray peaks were not accompanied by radio ejections (Tavani et al. 1996). The source then returned to a quiescent state from late 1995 until early 1996 when a new outburst took place. The then new {\sl RXTE} monitored the 2--10 keV X-rays; the lightcurve rather surprisingly reached a plateau while the optical decayed, contrary to the more correlated behaviour expected (Orosz et al. 1997, Hynes et al. 1998). 

Since this last outburst, the source has remained in quiescence until 2005 February 17 when an increase in X-ray flux was detected by {\sl RXTE}/PCA (Markwardt \& Swank 2005). As the outburst evolved, the X-ray spectrum was seen to pass through various X-ray spectral states (e.g. Markwardt et al. 2005, Homan 2005) and a variable radio source was detected, initially with a flat spectrum as expected for a jet associated with the low/hard state (e.g. Rupen et al. 2005a, b, c; Brocksopp et al. in prep.). 

GRO~J1655$-$40 is the first black hole X-ray transient to have been monitored by {\sl Swift}. In this paper we present results from all three of the {\sl Swift} instruments, both individually and combined together as a single broad-band dataset.

\begin{table*}
\caption{Details of the XRT observations. The columns list the observation ID number, the associated modified Julian date, the start date and time of the XRT observations, the net exposure time, the observing mode and the background-subtracted count rate detected in the 0.7--9.6
 keV range (with standard deviation in brackets).}
\begin{tabular}{lccccc}
\hline
\hline
Observation ID &MJD &Start Date (Time) & Exposure (s) & Mode & Count Rate (0.7--9.9 keV)\\
\hline
00030009002 &53435	 &2005-03-06 (10:01:14) & 2709& WT	& 18.6        (0.1)\\
00030009005 &53448	 &2005-03-19 (01:47:34) & 1711& LrPD	& 1041.0      (0.8)  \\    
00030009006 &53449	 &2005-03-20 (03:42:37) & 3   & LrPD	& 1044.0       (19.6)  \\    
00030009007 &53450	 &2005-03-21 (05:11:54) & 1397& LrPD	& 911.9      (0.9)      \\
00030009008 &53456	 &2005-03-27 (09:03:58) & 2836& LrPD	& 625.9      (0.5)      \\
00030009011 &53470	 &2005-04-10 (10:35:14) & 921 & LrPD	& 807.1      (1.0)      \\
00030009012 &53481	 &2005-04-21 (21:31:34) & 218 & LrPD	& 925.6       (2.1)      \\
00030009014 &53493	 &2005-05-03 (23:15:59) & 1143& LrPD	& 944.1      (1.0)      \\
00030009015 &53504	 &2005-05-14 (08:01:48) & 1990& LrPD  & 1217.0      (0.8)      \\
00030009016 &53505	 &2005-05-15 (10:06:55) & 911 & LrPD	& 1155.0       (1.2)      \\
00030009017 &53506	 &2005-05-16 (11:29:34) & 1946& LrPD	& 1678.0      (1.0)      \\
00030009018 &53511	 &2005-05-21 (08:52:32) & 1915& LrPD	& 1336.0      (0.9)      \\
00030009019 &53512	 &2005-05-22 (07:15:55) & 2346& LrPD	& 1098.0      (0.7)      \\
00030009020 &53513	 &2005-05-22 (20:34:54) & 507 & LrPD	& 1138.0       (1.5)      \\
00030009021 &53525       &2005-06-04 (03:55:11) & 697 & WT    & 689.5       (1.0)      \\
00030009022 &53526       &2005-06-05 (02:24:55) & 702 & WT    & 677.8       (1.2)      \\
00030009023 &53527       &2005-06-06 (02:34:56) & 639 & WT    & 635.0       (1.3)      \\
00030009025 &53540       &2005-06-19 (02:29:43) & 909 & WT    & 490.9      (0.8)      \\
00030009026 &53545       &2005-06-23 (19:28:32) & 328 & WT    & 406.6       (1.1)      \\
\hline 
\label{tab:xrtobs}
\end{tabular}
\end{table*}

\begin{table*}
\caption{UVOT Photometry in the six filters; units are magnitudes, errors are $1\sigma$ statistical errors (determined from the number of counts in the sky aperture). Central wavelengths for the filters are: $V\, (5460\, \mbox{\AA})$, $B\, (4350\, \mbox{\AA})$, $U\, (3450\, \mbox{\AA})$, $UVW1\, (2600\, \mbox{\AA})$, $UVM2\, (2200\, \mbox{\AA})$ and $UVW2\, (1930\, \mbox{\AA})$. $UV$ ``magnitudes'' are defined in the usual way as $-2.5\,\log\,(\mbox{counts})$ for consistency.}
\begin{tabular}{lccccccc}
\hline
\hline
Observation ID	&MJD 	   & 	 V 		& B 		& U 		& UVW1 		& UVM2 	& UVW2\\
\hline
00030009002&	53435 & 17.11 (0.11)   &  18.53 (0.12) 	& 18.87 (0.21) 	 & 20.67 (0.43)  & 21.31 (0.50)    &22.05 (0.67)  \\
00030009005&	53448 & 16.28 (0.10)   &  --		& --		 & 18.85 (0.32)  & 22.09 (0.92)    &20.64 (0.35)  \\
00030009006&	53449 & 15.90 (0.10)   &  --		& --		 & 18.34 (0.41)  & 20.61 (0.41)    &20.44 (0.34)  \\
00030009008&	53456 & 15.56 (0.10)   &  --		& --		 & 18.30 (0.27)  & 20.08 (0.42)    &19.81 (0.30)  \\
00030009011&	53470 & 15.32 (0.09)   &  16.76 (0.09) 	& 16.73 (0.17) 	 & 18.32 (0.25)  & 20.60 (0.40)    &19.50 (0.29)  \\
00030009012&	53481 & --	       &  --		& --		 & --		 & 19.63 (0.39)    &--		    \\
00030009015&    53504 & 16.10 (0.10)   &  17.66 (0.11)  & 17.61 (0.17)   & 19.09 (0.27)  & 21.40 (0.60)    &20.74 (0.37)  \\
00030009017&	53506 & 16.52 (0.10)   &  18.09 (0.11)  & 18.35 (0.19)   & 19.76 (0.31)  & 21.58 (0.69)    &20.69 (0.36)  \\
00030009018&	53511 & 16.67 (0.11)   &  18.18 (0.12)  & 18.33 (0.19)   & 20.27 (0.32)  & --		   &20.84 (0.39)  \\
00030009019&	53512 & 16.26 (0.10)   &  18.37 (0.39)  & 17.92 (0.30)   & 21.65 (0.40)  & --		   &20.87 (0.38)  \\
00030009021&	53525 & 16.40 (0.10)   &  17.94 (0.10)  & 18.38 (0.19)   & 19.49 (0.29)  & 21.53 (0.68)    &21.10 (0.43)  \\ 
00030009023&	53527 & 16.80 (0.11)   &  18.34 (0.11)  & 18.71 (0.21)   & 20.60 (0.47)  & 21.42 (0.75)    &21.79 (0.69)  \\
00030009025&    53540 & 16.91 (0.11)   &  18.39 (0.11)  & 19.11 (0.23)   & 20.30 (0.38)  & 21.00 (0.48)    &22.37 (1.04)  \\
00030009026&    53545 & 16.65 (0.16)   &  --            &	--       & 	--	 &	--         &20.74 (0.67)  \\
\hline
\hline 
\label{tab:uvotobs}
\end{tabular}
\end{table*}

\section{Observations}
{\sl Swift} observed GRO~J1655$-$40 on 20 occasions between 2005 March 6 and June 23, in addition to more continuous monitoring by the BAT. Most pointed observations obtained data with all three instruments, details of which are listed in Tables~\ref{tab:xrtobs},~\ref{tab:uvotobs}.~\ref{tab:batmodel},~and~\ref{tab:xrtmodel}.

\subsection{Burst Alert Telescope}

The BAT data were reduced using the standard {\it Swift} software and the resultant spectra were fit using {\sc XSPEC} (Arnaud et al. 1996). The BAT dataset for GRO J1655-40 is more extensive than that of the XRT or UVOT since the BAT can observe the source when it is significantly off-axis. BAT data were accumulated for all observations in which GRO~J1655$-$40 was sufficiently close to the field centre that it illuminated at least $\sim10$\% of the BAT detectors.  This corresponds to $\sim38^{\circ}$ from the centre in the short direction and $\sim55^{\circ}$ in the long direction of the approximately rectangular BAT field of view. The count-rate was determined from the sky images by subtracting an annular background and fitting the BAT point spread function (PSF). The BAT PSF is approximated by a two-dimensional pyramidal frustum with a square base and full width half maximum 22.45\arcmin (the BAT angular resolution) and top width 2.75\arcmin. The background size is constant in the sky image tangent plane and, at the field centre, has diameters of $1.6^{\circ}$ (inner) and $7.4^{\circ}$ (outer); this restricts the background to a local region of the sky while eliminating source counts. 

Before performing spectral analysis a number of corrections were made to the raw BAT survey detector plane histograms. Firstly, the histograms were rebinned in energy to correct a small residual inaccuracy, arising from the way the histograms are created on-board the instrument. Secondly, noisy detectors with a count rate more than $\sim$twice the mean rate were removed from the analysis dataset. Thirdly, the background and other bright sources in the large BAT field of view were removed as described in the remainder of the paragraph. The BAT background varies with position in the orbit, but averages $\sim8000$ cts/sec (15--350 keV) for the whole instrument. In addition the low-mass X-ray binary Sco X-1 was in the field for most observations at a flux rate (15--25 keV) $\sim3$ times that of GRO~J1655$-$40. In some observations the high-mass X-ray binary GX 301-02 was also cleaned. The most effective subtraction method is mask-tagging which uses the BAT tool {\sc batmaskwtimg} to encode every detector with a weighting factor, corresponding to the fraction of the detector which was exposed to the source through the empty cells in the BAT coded aperture. A detector fully open to the source receives a weight of +1 and a fully blocked detector a weight of $-1$.  When the BAT histogram is convolved with this weighting matrix, nearly all of the contributions from background and extraneous sources are removed. The remaining contamination ($\sim$2--10 \% for this data set) is subtracted by ``cleaning''. A standard background model, that includes variation across the BAT detector array, is fit to the data and then removed. Similarly, a projection of each bright source through the coded aperture is fit to the focal plane and removed in a single iteration. Finally, all detectors lying on the projection of bright sources through the edge of the coded aperture were removed from the analysis data set. This last step is necessary because the geometry and absorption of gaps in the coded aperture and graded-Z shield near the edges are not well-known and hence cannot be well-modelled in the response function. The response function was generated for each observation using the {\it Swift} {\sc batdrmgen} routine. It correctly accounts for disabled and ignored detectors and geometric effects when GRO~J1655$-$40 is not in the centre of the BAT field of view.

\subsection{X-ray Telescope}

The XRT data were reduced using version 2.0 of the standard {\it Swift} software.  A number of tasks were performed by employing the perl script {\sc xrtpipeline}, using the default parameters as described in the XRT users' manual\footnote{http://www.swift.ac.uk/proc\_guide.shtml}. The processing steps include identifying bad pixels or bad columns, performing coordinate transformations, time-tagging and  reconstruction of events, computation of the pulse height analysis and pulse invariant values and elimination of frames affected by partially exposed pixels. The data were processed using the default good-time-interval screening parameters, the criteria for which are grade-selection plus the removal of calibration sources, bad pixels (dead, hot and flickering), pixels below the event threshold for Earth-limb screening, saturated pixels, partially exposed pixels and telemetry-saturated frames (again, see the XRT users' manual for additional information). 

The XRT observations took place in one of two observing modes which have been optimised for gamma-ray burst follow-up and rapid telemetry. The Low-Rate Photodiode mode (LrPD; Burrows et al. 2005, Hill et al. 2004) is used for the brightest sources and has a time resolution of $10\mu$sec. This mode provides no spatial information and, instead, the image is condensed to a single pixel, from which the spectrum of the source is extracted. Background-subtraction is therefore currently only possible by extracting a spectrum from the slew data preceeding each pointed observation. While this provides adequate subtraction of the background, there are residuals in the regions of 5.5--7.0 keV due to the iron calibration sources in the camera. We therefore urge caution at over-interpretation of any narrow-band spectral features in this range at the current time. The Windowed Timing (WT; Burrows et al. 2005, Hill et al. 2004) mode has a time resolution of 1.8 msec and provides one-dimensional spatial information. It is therefore possible to extract a background spectrum from the one-dimensional image without leaving residuals in the iron-line region. An extraction window of 20 pixels (47\arcsec) was used for the source-spectrum, with an additional extraction-window of 35 pixels on either side of the source for the background. The background regions were positioned either side of the source so as to ensure an evenly-sampled background. We note that the WT mode was used for some of the later observations due to damage to the CCD detector by a micro-meterorite, resulting in the loss of the LrPD mode (Abbey, Stevenson \& Ambrosi 2005). Consequently, comparison of these later WT data with the expected one-dimensional WT PSF (Moretti et al. 2005) showed that the central part of the PSF was affected by pile-up. In order to correct for this effect, we excluded data within a box of length 4 pixels (9.45 \arcsec) centered on the centroid of the (1-D) PSF.

In all cases, the spectra were extracted from the events files using the {\sc xselect} software, binned to 30 counts using {\sc grppha} and loaded into {\sc xspec} (Arnaud 1996) for model-fitting. We used version 7 of the response matrices and created individual ancillary response files (ARF) using {\sc xrtmkarf}. In the case of the piled-up spectra, the PSF correction is applied directly to the ARF, again by {\sc xrtmkarf}. The energy-resolution of the XRT is FWHM$\sim142$ eV at 5.96 keV.

\subsection{Ultraviolet/Optical Telescope}

Several observations of GRO J1655-40 were taken with the UVOT in the six different filters - $V\, (5460\, \mbox{\AA})$, $B\, (4350\, \mbox{\AA})$, $U\, (3450\, \mbox{\AA})$, $UVW1\, (2600\, \mbox{\AA})$, $UVM2\, (2200\, \mbox{\AA})$ and $UVW2\, (1930\, \mbox{\AA})$. Again, the data were processed using the standard {\sl Swift} software. Some data were taken in ``event-mode'' in which the time and detector position of each photon are recorded; others were taken in ``imaging-mode'', in which the image is accumulated on-board to conserve telemetry volume, discarding the photon timing information within an exposure. Event-mode data were screened for standard bad times e.g. South Atlantic Anomaly passage, Earth-limb avoidance. These files contain values for corrected detector and sky coordinates. Images were extracted from the event mode data using {\sc xselect}. Image-mode data were screened for bad times, corrected for Mod-8 noise and had their sky coordinates determined. For both observing modes the coordinates of the resulting images were corrected for the $\sim5$\arcsec uncertainty in the aspect of the spacecraft pointing. A number of images were taken with UVOT with typical exposure times ranging from 2 to 200 seconds for the optical filters and 70 to 700 seconds for the UV filters. In order to improve the signal to noise ratio, we have summed each set of observations for each filter using the {\sl Swift} task {\sc uvotimsum}. Aperture photometry was performed on the images using {\sc GAIA} (the Graphical Astronomy and Image Analysis tool produced by the Central Laboratory of the Research Councils, UK) using an aperture of $\sim 2.5 \arcsec$ radius. Background subtraction was performed using counts extracted from a larger region (2--3 times the radius of the source-aperture) off-set from the source position. The measured counts per second were converted to magnitudes using the preliminary zero-points, measured in orbit.

\section{Lightcurves}

\begin{figure*}
\begin{center}
\leavevmode
\epsfig{file=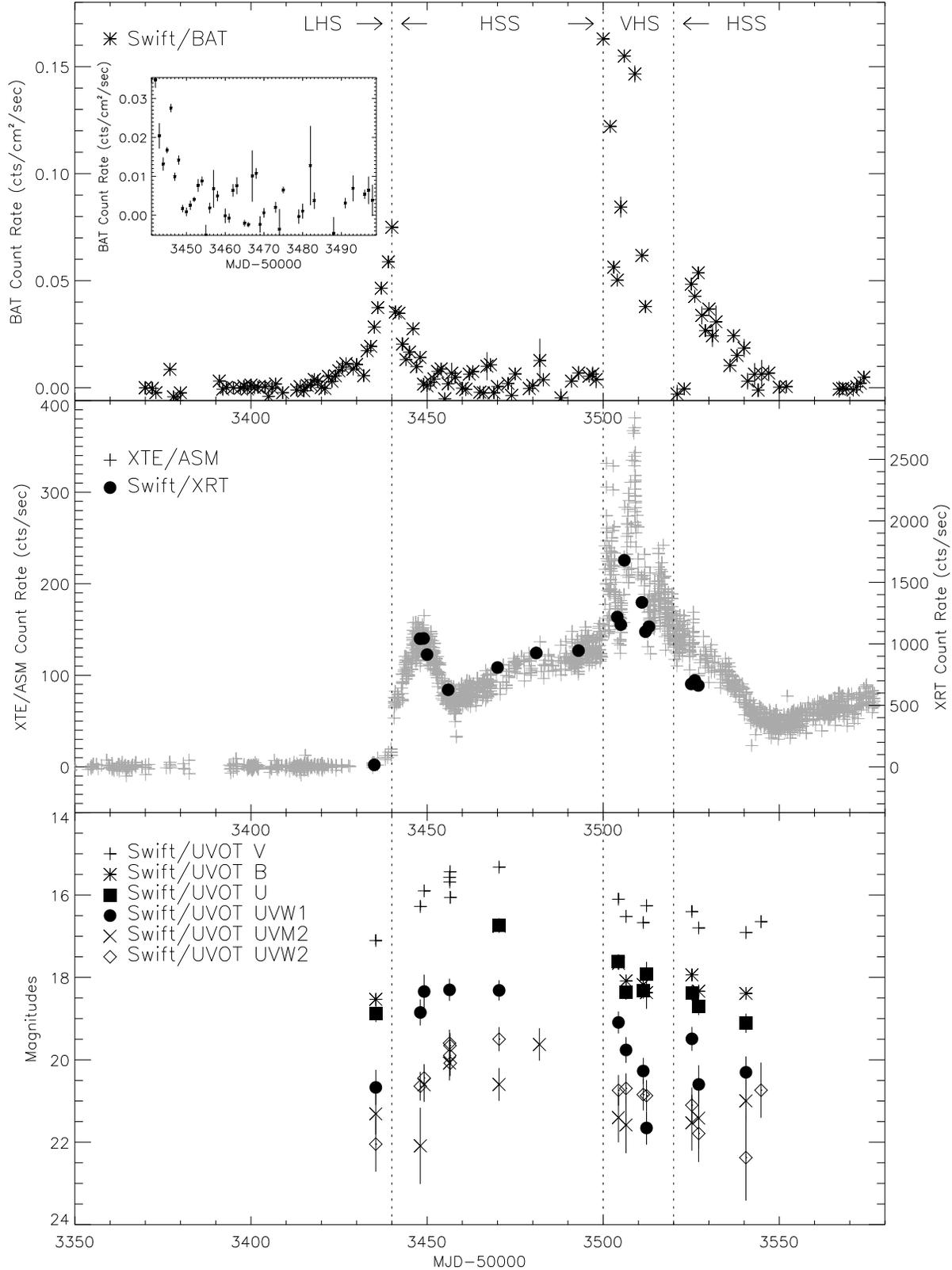, width=17cm}
\vspace*{-1cm}
\caption{Lightcurves showing the 2005 outburst of GRO~J1655$-$40. The BAT (15--50 keV) data is plotted in the top panel, with an inset panel enlarging the period MJD 53449--53499. The middle panel shows the {\sl RXTE}/ASM data (+) with the XRT pointed observations overplotted ($\bullet$). The UVOT photometry is plotted in the bottom panel, the filters indicated as shown in the key. We discuss the high levels of variability and correlation in the text. Approximate durations of each spectral state are indicated (LHS=Low/hard state; HSS=High/soft state; VHS= Very high state).}
\label{fig:lightcurves}
\end{center}
\end{figure*}

Lightcurves for the three {\sl Swift} instruments are plotted in Fig.~\ref{fig:lightcurves}, along with the public {\sl RXTE}/ASM lightcurve for comparison. The X-ray points are measured in count-rates and the optical/ultraviolet in magnitudes. All three plots clearly show significant and complex variabilty; in particular both X-ray plots display a sequence of peaks taking place on various timescales.

\begin{figure*}
\begin{center}
\leavevmode
\epsfig{file=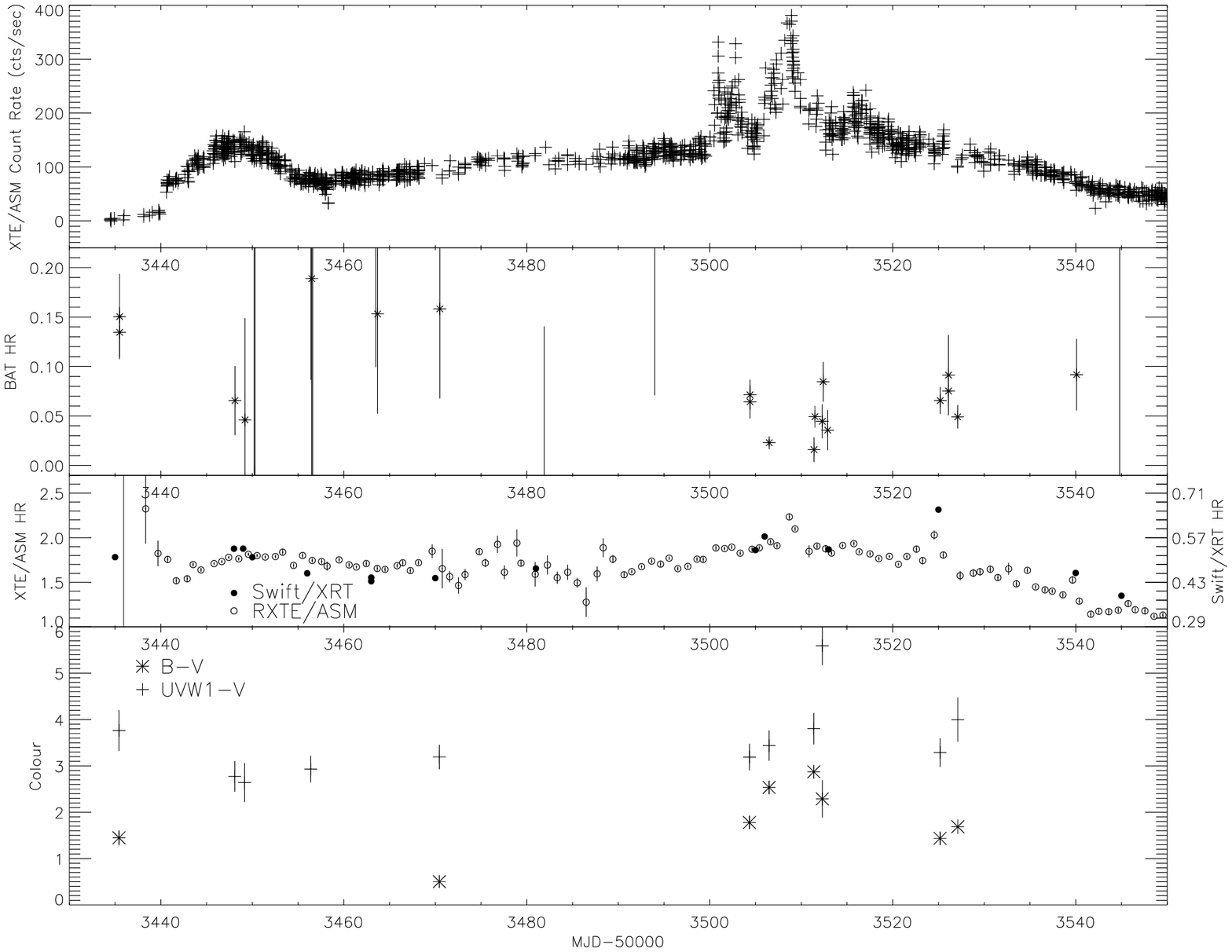, width=12cm,angle=90}
\caption{The {\sl RXTE}/ASM lightcurve plotted with the 100--150 keV / 14--25 keV (second panel down), 0.5--3 keV / 3--9.5 keV (third panel down, solid points) and 3--10 keV / 1.5--3 keV (third panel down, open points) hardness ratios for the BAT, XRT and ASM respectively. The bottom panel shows the $B-V$ and $UVW1-V$ optical/ultraviolet colours.}
\label{fig:hardness}
\end{center}
\end{figure*}

Despite the outburst alert coming from soft (2--10 keV) X-ray observations using the {\sl RXTE}/PCA, the source initially became active in the hard X-rays (14--150 keV), rising to a sharp peak. It then decayed and entered a ``plateau phase'', during which there was continued low-level variability for approximately one month (inset plot in Fig.~\ref{fig:lightcurves}). The source then rose steeply and entered a period of dramatic variability for $\sim 3$ weeks, reaching a maximum flux of $\sim$ (2--3) $\times 10^{-8}\mbox{erg}\,\mbox{cm}^{-2}\,\mbox{s}^{-1}$. While initial inspection might suggest similar behaviour in the soft X-rays, closer investigation reveals this not to be the case. The first peak of the soft X-ray lightcurve was delayed by approximately one week relative to the hard X-rays. It then decayed steeply and, rather than enter a plateau phase, started a long period of gradual brightening. The soft X-ray source also then entered a much brighter and more variable phase; again, the peak initially took place in the hard X-rays but this time with a delay of approximately one day. The time-sampling of the {\it Swift} observations does not allow detailed study of the variable phase but the BAT and ASM data show that at least three peaks occured.

The optical and ultraviolet data show very different behaviour from the X-ray. While of a lower time-sampling, it nonetheless appears that the source rose slowly from quiescence (the first $V$-band point is approximately equivalent to quiescence) by nearly 2 magnitudes in the $V$-band. The gradient remained approximately constant, with no deviation despite the rise and decay of the initial X-ray peak during this period. A peak of $V\sim$ 15.3 magnitudes was reached and ROTSE data obtained by Smith (2005) showed a flat optical lightcurve during the following period of gradually-rising soft X-ray flux. Once the X-ray lightcurves entered the more variable phase, there was a marked drop in the optical/ultraviolet flux, as also shown by Smith (2005) and in the infrared by Buxton \& Bailyn (2005).

The three phases of the lightcurve (initial peak, gradual rise, bright/variable) have already been identified in the literature as being associated with the ``canonical'' low/hard, high/soft and very high states of black hole transients (Homan 2005; Homan et al. 2005). The dates of the respective transitions are approximately March 11 (MJD 53440) for the LHS--HSS and May 11 (MJD 53501) for the LHS--VHS, as indicated in Fig.~\ref{fig:lightcurves} by vertical dashed lines.

We replot the {\sl RXTE}/ASM lightcurve in the top panel of Fig.~\ref{fig:hardness} for comparison with the X-ray hardness ratios and optical/ultraviolet colours. The second panel shows the BAT hardness ratio (100--150 keV / 14--25 keV), the third panel shows the {\sl RXTE}/ASM (3--10 keV / 1.5--3 keV) and XRT (3--9.5 keV / 0.5 -- 3 keV) hardness ratios. Two optical/UV colours are plotted in the bottom panel. It is clear, from the ASM hardness ratio, that the spectrum was initially hard and then softened at the beginning of the first low energy X-ray peak. It then hardened slightly before remaining approximately constant until the period of high variability. During this ``very high state'' there appears to be a phase of ``temporary hardening'' in the ASM hardness ratio. Similarly there appears to be an additional temporary hardening on MJD 53525 which features in the ASM and XRT hardness ratios; there is no obvious corresponding feature in the 2--10 keV lightcurve but Fig.~\ref{fig:lightcurves} shows that there {\em was} a hard X-ray flare at this time. The signal-to-noise of the BAT hardness ratio renders any variability less convincing, but there is a clear BAT softening during the initial soft X-ray peak, coincident with the onset of the high/soft state. There also appears to be a softening of the BAT data 2--3 days {\em prior} to the the largest soft X-ray flare, which took place on MJD 53508. There is additional possible variability in the ASM hardness ratio during the high/soft state which may be related to the small-scale variability observed by the BAT at this time. Finally the two low energy X-ray hardness ratio plots show a gradual softening following the return to the high/soft state and gradual decay of the X-ray flux.

\begin{table*}
\caption{Results of fitting a simple power-law model to those BAT spectra bright enough for the fit to be constrained. The first two columns list the Observation ID and the date. The next two columns list the power-law photon index and normalisation (in photons/keV/cm$^2$/s at 1 keV) values. The following two columns list fluxes in specific BAT energy ranges (values $\times10^{-10}\mbox{ergs cm}^{-2}\mbox{s}^{-1}$) and the final three columns list the values of $\chi^2$, reduced $\chi^2$ and the number of degrees of freedom. All errors and upper limits are to within 90\% confidence. From the low values of $\chi^2_{\nu}$ and the power-law normalisation, it is clear that the signal-to-noise is low at some epochs. The low values of $\chi^2_{\nu}$ are due to the inclusion of the systematic error vector, as detailed on http://swift.gsfc.nasa.gov/docs/swift/analysis/bat\_digest.html}
\begin{tabular}{lcccccccc}
\hline
\hline
Obs. ID &MJD&Photon&PL &15-25 keV& 15-350 keV&$\chi^2$&$\chi^2_{\nu}$&D.O.F\\
&&Index&Norm.&Flux&Flux&&&\\
\hline

9002a  &  53435.5	&   $1.55^{+0.10}_{-0.10}$ & $0.19^{+0.09}_{-0.06}$    &    $5.8^{+0.7}_{-0.6}$ & $ 70.0^{+6.7}_{-6.4}	   $ &  22.01 & 0.37 & 59\\
9002b  &  53435.5	&   $1.55^{+0.13}_{-0.13}$ & $0.17^{+0.11}_{-0.07}$    &    $5.3^{+0.8}_{-0.7}$ & $ 65.0^{+9.5}_{-8.1}	   $ &  35.97 & 0.61 & 59\\
9005a  &  53448.1	&   $1.93^{+0.18}_{-0.18}$ & $0.44^{+0.41}_{-0.21}$    &    $4.3^{+0.8}_{-0.7}$ & $ 29.7^{+5.0}_{-4.8}	   $ &  34.41 & 0.58 & 59\\
9006a  &  53449.2	&   $1.83^{+0.40}_{-0.38}$ & $0.15^{+0.44}_{-0.15}$    &    $2.0^{+0.7}_{-0.6}$ & $ 15.1^{+7.2}_{-5.0}	   $ &  36.37 & 0.62 & 59\\
9007a  &  53450.2	&   $0.79^{+1.98}_{-3.79}$ & $<0.70$                   &    $<2.5$              & $ 17.7^{+98.9}_{-15.4}   $   &  35.01 & 0.59 & 59\\
9007c  &  53450.3	&   $3.29^{+1.78}_{-1.26}$ & $<1.9\times10^3$          &    $2.6^{+1.1}_{-1.7}$ & $ 5.9^{+5.9}_{-3.8}	   $ &  28.99 & 0.49 & 59\\
9008a  &  53456.4	&   $5.53^{+4.47}_{-3.67}$ & $<2.7\times10^8$          &    $1.1^{+0.5}_{-1.0}$ & $ 1.4^{+4.3}_{-1.1}	   $ &  42.14 & 0.71 & 59\\
9008c  &  53456.5	&   $0.59^{+9.41}_{-3.59}$ & $<1.7\times10^7$          &    $<0.7$              & $ 0.4^{+23.4}_{-0.2}	   $ &  44.67 & 0.76 & 59\\
9008d  &  53456.6	&   $9.10^{+0.90}_{-8.16}$ & $<1.7\times10^7$          &    $<0.7$              & $ 0.4^{+30.8}_{-0.2}	   $ &  37.70 & 0.64 & 59\\
9010a  &  53463.7	&   $7.48^{+2.52}_{-8.07}$ & $<3.5\times10^7$          &    $1.1^{+0.5}_{-0.6}$ & $ 1.1^{+0.9}_{-0.6}	   $ &  35.44 & 0.60 & 59\\
9011a  &  53470.5	&   $4.21^{+5.79}_{-2.24}$ & $<1.3\times10^7$          &    $1.5^{+0.6}_{-0.9}$ & $ 2.2^{+1.9}_{-1.1}	   $ &  33.32 & 0.56 & 59\\
9014a  &  53494.0	&   $0.53^{+9.47}_{-3.46}$ & $<1.0\times10^7$          &    $1.0^{+0.6}_{-0.9}$ & $ 1.4^{+5.1}_{-1.1}	   $	  &  42.67 & 0.72 & 59\\
9015a  &  53504.4	&   $2.30^{+0.14}_{-0.14}$ & $3.53^{+2.30}_{-1.38}$    &    $11.9^{+1.2}_{-1.2}$ & $ 51.4^{+4.7}_{-4.1}    $   &  36.05 & 0.61 & 59\\
9015b  &  53504.4	&   $2.28^{+0.16}_{-0.15}$ & $3.88^{+2.81}_{-1.61}$    &    $13.9^{+1.4}_{-1.3}$ & $ 61.4^{+6.5}_{-6.2}    $	  &  39.96 & 0.68 & 59\\
9017a  &  53506.5	&   $2.73^{+0.08}_{-0.08}$ & $46.11^{+15.82}_{-11.67}$ &    $43.5^{+2.7}_{-2.5}$ & $ 125.9^{+4.1}_{-4.4}   $	  &  41.13 & 0.70 & 59\\
9018a  &  53511.4	&   $2.38^{+0.13}_{-0.12}$ & $6.20^{+3.56}_{-2.24}$    &    $16.6^{+1.4}_{-1.4}$ & $ 65.8^{+4.9}_{-4.6}    $   &  28.00 & 0.47 & 59\\
9018b  &  53511.5	&   $2.37^{+0.11}_{-0.11}$ & $6.17^{+3.09}_{-2.04}$    &    $16.7^{+1.5}_{-1.4}$ & $ 66.6^{+4.4}_{-4.6}    $	  &  29.89 & 0.51 & 59\\
9019a  &  53512.3	&   $2.17^{+0.15}_{-0.15}$ & $2.00^{+1.43}_{-0.83}$    &    $10.0^{+1.2}_{-1.0}$ & $ 49.9^{+5.5}_{-5.0}    $	  &  38.11 & 0.65 & 59\\
9019b  &  53512.4	&   $2.15^{+0.18}_{-0.17}$ & $1.58^{+1.40}_{-0.74}$    &    $8.4^{+1.1}_{-1.0}$  & $ 43.3^{+5.2}_{-5.3}    $	  &  52.01 & 0.88 & 59\\
9020a  &  53512.9	&   $2.35^{+0.18}_{-0.17}$ & $3.43^{+2.98}_{-1.57}$    &    $10.0^{+1.3}_{-1.2}$ & $ 40.8^{+5.8}_{-4.7}    $	  &  37.92 & 0.64 & 59\\
9021a  &  53525.2	&   $2.06^{+0.10}_{-0.10}$ & $1.62^{+0.72}_{-0.50}$    &    $11.2^{+1.0}_{-1.0}$ & $ 64.3^{+4.8}_{-4.2}    $	  &  23.19 & 0.39 & 59\\
9022a  &  53526.1	&   $2.29^{+0.31}_{-0.28}$ & $2.93^{+5.32}_{-1.84}$    &    $10.3^{+1.7}_{-1.8}$ & $ 45.0^{+11.7}_{-8.5}   $	  &  47.79 & 0.81 & 59\\
9022b  &  53526.1	&   $2.10^{+0.13}_{-0.13}$ & $1.90^{+1.19}_{-0.73}$    &    $11.6^{+1.2}_{-1.2}$ & $ 62.6^{+5.5}_{-5.6}    $   &  36.51 & 0.62 & 59\\
9023a  &  53527.1	&   $2.25^{+0.11}_{-0.10}$ & $3.38^{+1.57}_{-1.07}$    &    $13.3^{+1.1}_{-1.1}$ & $ 61.0^{+4.1}_{-3.5}	   $ &  24.99 & 0.42 & 59\\
9025a  &  53540.1	&   $1.91^{+0.21}_{-0.21}$ & $0.32^{+0.37}_{-0.17}$    &    $3.5^{+0.6}_{-0.6}$  & $ 24.4^{+4.5}_{-3.6}    $	  &  37.32 & 0.63 & 59\\		  	   

\hline 
\label{tab:batmodel}
\end{tabular}
\end{table*}

\begin{table*}
\caption{Results of fitting simple spectral models to the XRT data. We restrict the fits to 0.7--9.6 keV because the spectral response outside this range is not yet completely optimized and is presently under investigation. However, we note that calibration uncertainties in the high energy region of the spectrum in some cases resulted in ``negative power-law slopes''; we therefore further restrict the upper energy limit, resulting in fits from 0.7 to 7.5 or 8.0 keV for these observations (*) and find that the power-law parameters are less well-constrained. The first observation (in the low/hard state) is fit with an absorbed power-law, plus a Gaussian of fixed width 0.01 keV; subsequent observations (in the high/soft or very high states) are fit with an absorbed power-law plus a multicolour disk blackbody (diskbb in {\sc xspec}). All errors and upper limits are to within 90\% confidence. An iron line is detected in the low/hard state observation, positioned at $6.39^{+0.05}_{-0.07}$ keV, with equivalent width $69^{+34}_{-36}$ eV (to 90\% confidence) and improving the fit by $4.6\sigma$ compared with an absorbed power-law alone.}
\begin{tabular}{lccccccccc}
\hline
\hline
Observation ID &Model 	& n$_{H}$ ($\times10^{22}\mbox{cm}^{-2}$) 	&Photon Index&PL Norm. &Disc Temp. (keV)&BB Norm.&$\chi^2$&$\chi^2_{\nu}$&D.O.F\\
\hline
00030009002 &PL + Gauss	&0.59$^{+0.01}_{-0.02}$ & 1.72$^{+0.03}_{-0.03}$     &0.25$^{+0.01}_{-0.01}$ &	 -- 		       & --              &667.2	 &1.17	& 571\\
00030009005 &PL + BB	&0.68$^{+0.01}_{-0.01}$ & 1.66$^{+0.13}_{-0.15}$     &*2.82$^{+0.69}_{-0.67}$ & 1.35$^{+0.01}_{-0.02}$  &725$^{+59}_{-58}$ &1255.3  &1.73  & 727\\
00030009007 &PL + BB	&0.69$^{+0.02}_{-0.02}$ & 2.12$^{+0.24}_{-0.20}$     &4.27$^{+0.88}_{-0.71}$ & 1.43$^{+0.04}_{-0.04}$  &492$^{+48}_{-43}$  &696.8	 &0.83 	& 841\\
00030009008 &PL + BB	&0.78$^{+0.02}_{-0.02}$ & 3.29$^{+0.20}_{-0.22}$     &7.65$^{+0.88}_{-0.75}$ &1.56$^{+0.01}_{-0.01}$   &245$^{+15}_{-15}$  &938.0	 &1.11  & 844\\
00030009011 &PL + BB	&0.68$^{+0.03}_{-0.02}$ & 2.48$^{+0.24}_{-0.16}$     &6.08$^{+1.15}_{-0.60}$ &1.42$^{+0.04}_{-0.04}$   &405$^{+27}_{-28}$  &799.0	 &1.03  & 773\\
00030009012 &PL + BB	&0.66$^{+0.03}_{-0.03}$ & 2.13$^{+0.20}_{-0.21}$     &5.51$^{+1.58}_{-1.24}$ &1.38$^{+0.06}_{-0.06}$   &503$^{+82}_{-89}$  &803.8	 &1.20  & 671\\
00030009014 &PL + BB	&0.66$^{+0.03}_{-0.02}$ & 2.21$^{+0.26}_{-0.18}$     &4.46$^{+1.36}_{-0.56}$ &1.48$^{+0.04}_{-0.04}$   &448$^{+29}_{-37}$  &1099.4 &1.33  & 827\\
00030009015 &PL + BB	&--                    & --                          &--  	             &--		       &--                &2193.7 & 2.41	& 910\\
00030009016 &PL + BB	&0.60$^{+0.02}_{-0.01}$ & $1.57^{+0.13}_{-0.14}$     &*3.82$^{+0.89}_{-0.83}$ &1.38$^{+0.04}_{-0.03}$   &628$^{+80}_{-80}$ &932.4 & 1.28 & 727\\
00030009017 &PL + BB	&0.63$^{+0.01}_{-0.01}$ & 0.88$^{+0.16}_{-0.32}$     &2.41$^{+1.38}_{-1.69}$ &1.21$^{+0.03}_{-0.01}$   &1618$^{+33}_{-114}$ &1287.7 & 1.40 & 917\\
00030009018 &PL + BB	&0.63$^{+0.01}_{-0.01}$ & $1.58^{+0.08}_{-0.08}$     &*5.02$^{+0.69}_{-0.62}$ &1.33$^{+0.02}_{-0.02}$   &807$^{+62}_{-68}$ &1300.6 & 1.79 & 727\\
00030009019 &PL + BB	&0.64$^{+0.01}_{-0.01}$ & 1.71$^{+0.08}_{-0.07}$     &4.17$^{+0.67}_{-0.33}$ &1.38$^{+0.02}_{-0.02}$   &617$^{+25}_{-59}$  &1375.7 & 1.51 & 909\\
00030009020 &PL + BB	&0.67$^{+0.03}_{-0.03}$ & $2.07^{+0.36}_{-0.26}$     &*4.23$^{+1.32}_{-1.15}$&1.49$^{+0.05}_{-0.05}$    &547$^{+74}_{-53}$  &791.8  & 1.10 & 720\\
00030009021 &PL + BB	&$\le0.79$              & $\le0.77$                  &*0.43$^{+0.93}_{-0.23}$&1.12$^{+0.20}_{-0.02}$    &820$^{+84}_{-158}$     &1278 & 1.89 & 677\\
00030009022 &PL + BB	&0.78$^{+0.01}_{-0.06}$ & $\le0.79$                  &*0.76$^{+0.77}_{-0.63}$&1.12$^{+0.12}_{-0.05}$    &1176$^{+107}_{-259}$  &1024.7 & 1.51 & 677\\
00030009023 &PL + BB	&0.84$^{+0.01}_{-0.03}$ & $\le0.84$                  &*0.95$^{+0.10}_{-0.82}$&1.09$^{+0.12}_{-0.04}$    &1124$^{+111}_{-282}$  &912.2  & 1.25 & 727 \\
00030009025 &PL + BB	&0.73$^{+0.01}_{-0.01}$ & 0.52$^{+0.16}_{-0.40}$     &0.29$^{+0.39}_{-0.17}$&1.05$^{+0.04}_{-0.03}$    &1135$^{+78}_{-93}$ &1149.7 & 1.49 & 772\\
00030009026 &PL + BB	&0.75$^{+0.05}_{-0.03}$ & 1.71$^{+1.37}_{-0.63}$     &1.74$^{+1.34}_{-1.27}$&0.98$^{+0.02}_{-0.02}$    &1055$^{+169}_{-197}$ &747.5  & 1.24 & 602\\
\hline 
\label{tab:xrtmodel}
\end{tabular}
\end{table*}

As with the lightcurves, the UVOT colours seem to show little correlation with the X-ray plots. They appear to become slightly bluer at the time of the very high state but more points would be needed to be sure.

We investigate the spectral properties of the three states in more detail in Section 5.

\section{Correlations}

\begin{figure}
\begin{center}
\leavevmode
\epsfig{file=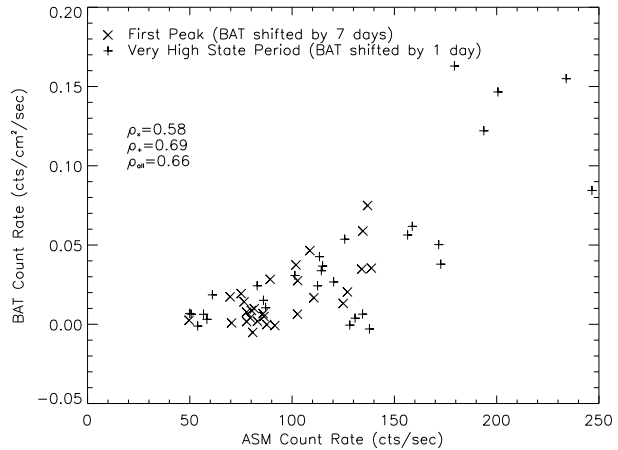,width=8cm}
\caption{BAT count rate plotted against the ASM count rate for the two sections of the lightcurves suggesting correlated behaviour. The BAT lightcurves were first shifted by seven days and one day for the two sections respectively to compensate for the soft X-ray delay. The first period is MJD 53440--53470 (initial peak), and the second is MJD 53498--53549 (very high state and its decay). The Spearman rank correlation coefficients ($\rho$) for these two periods are 0.58 and 0.69 respectively. When the two sections of data are combined, the overall correlation coefficient is 0.66. All values of $\rho$ are to $>99\%$ confidence.}
\label{correlate}
\end{center}

\end{figure}

In Section 2 we suggested that there was some degree of correlated behaviour between the lightcurves. In order to test this further, we have taken two sections of the X-ray (BAT and ASM) data. The first spans the time of the initial X-ray peak and the second spans the time of the very high state (and its subsequent decay). These periods correspond to MJD 53440--53470 and MJD 53498--53549 for the ASM data. The BAT lightcurve was shifted seven and one days for each section respectively to compensate for the soft X-ray delay. In Fig.~\ref{correlate} we plot the BAT count rate against the ASM count rate and calculate the Spearman rank order correlation coefficient ($\rho$) for these two sections of data, both individually and combined. Values of $\rho$ are listed on the plot and fall in the range 0.6--0.7 (to $>99\%$ confidence), suggesting a significant correlation at these times.

\section{Spectroscopy}

\subsection{Burst Alert Telescope}

Pointed BAT observations were obtained on 17 occasions, as well as the continued monitoring when GRO~J1655$-$40 was serendipitously in the large field of view. The BAT data were collected in ``survey mode'', in which events during a survey interval are accumulated in histograms with 80 energy channels. The time-resolution for these observations is the survey interval duration, which was typically either 300 sec or 450 sec. In order to improve statistics, all survey histograms for a given spacecraft pointing are added together before fitting. We fit the resultant spectra with a simple power-law in the 14--150 keV range. This model provides a reasonable fit to some of the spectra, although others were too faint for the statistics to be reliable. Interestingly, GRO J1655$-$40 is detected in the hard X-ray regime during each of the three X-ray spectral states. The photon indices were mainly in the range $\sim 2.0 - 2.4$ although we discuss this further in Section 5.3. The BAT energy resolution is $\sim5$ keV at 60 keV.

\subsection{X-ray Telescope}

\begin{figure}
\begin{center}
\leavevmode
\epsfig{file=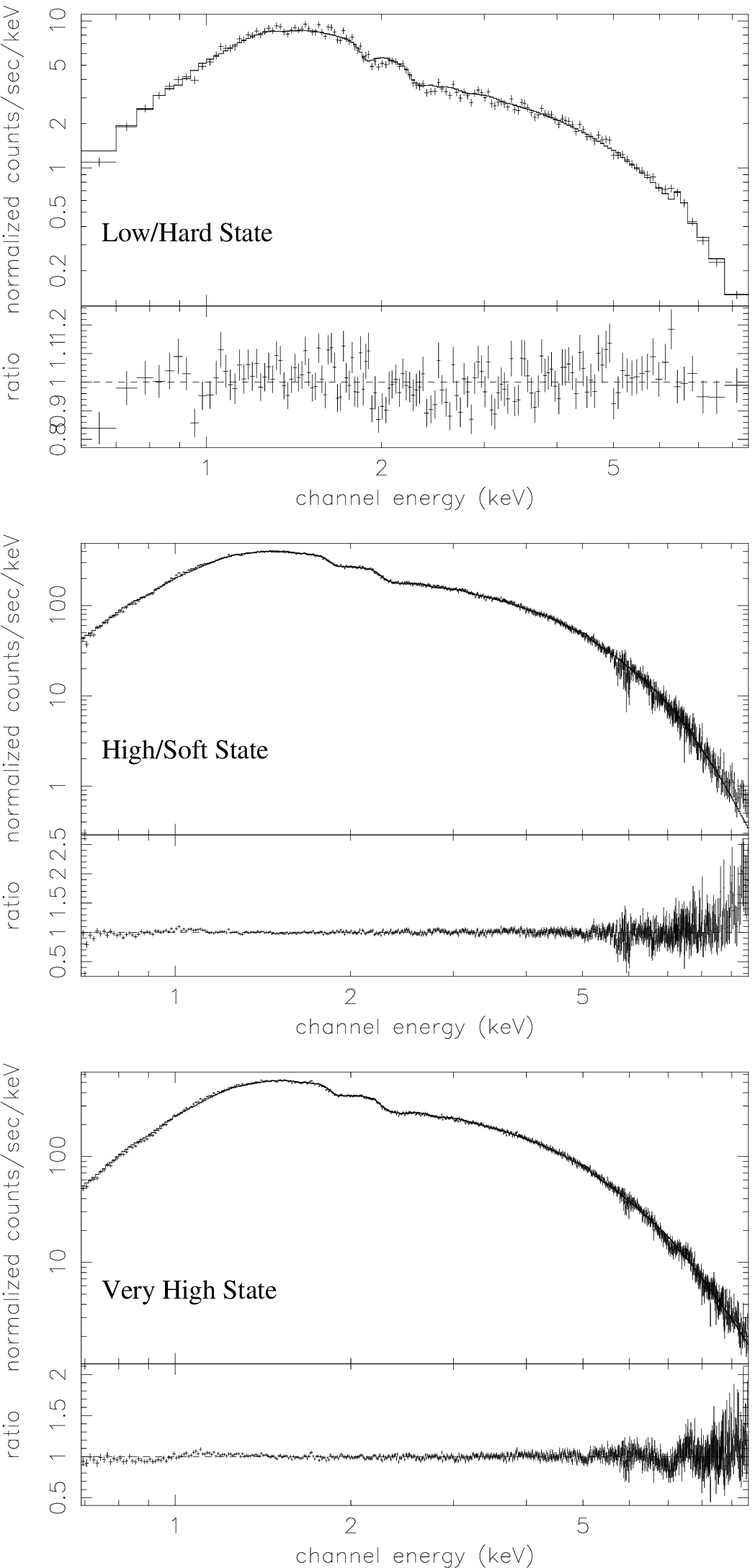,width=6cm}
\caption{Selected XRT spectra (ID 00030009002, 00030009011, 00030009016) showing the low/hard (top), high/soft (middle) and very high (bottom) states. The low/hard state spectrum has been fit with an absorbed power-law plus iron line of fixed width 0.01 keV. This iron line has energy $6.39^{+0.05}_{-0.07}$ keV, equivalent width $69^{+34}_{-36}$ eV (to within 90\% confidence) and improves the fit by $4.6\sigma$ compared with an absorbed power-law alone. The two softer spectra have been modelled with an absorbed power-law plus multicoloured disc blackbody (diskbb in {\sc xspec}). We emphasize that the ``features'' seen around 6--7 keV in the LrPD spectra are possible artefacts due to the calibration sources in this mode; further analysis of this region will be performed once improved background spectra can be obtained in the future. However, preliminary improvement of the background suggests that, during the high/soft and very high states, iron absorption lines are present at $6.782\pm0.031$ keV and/or $7.065\pm0.029$ keV, consistent with Ueda et al. 1998. Fit parameters are listed in Table~\ref{tab:xrtmodel}.} 
\label{fig:xrt-spectra}
\end{center}
\end{figure}

Fig.~\ref{fig:xrt-spectra} shows three of the XRT spectra, one for each of the three spectral states observed. The top panel shows the low/hard state spectrum from MJD 53435, the middle panel presents the high/soft state spectrum from MJD 53470 and the bottom panel presents the very high state spectrum from MJD 53505. The low/hard state spectrum has been fit with a simple absorbed power-law plus Gaussian; the other two spectra have been fit with an absorbed multicolour disc blackbody (diskbb in {\sc xspec}) in addition to the power-law. We note that these are standard phenomenological models for the respective spectral states, not a suggestion that these are necessarily the best physical models. Fit parameters are listed in Table~\ref{tab:xrtmodel}.

\begin{figure}
\begin{center}
\leavevmode
\epsfig{file=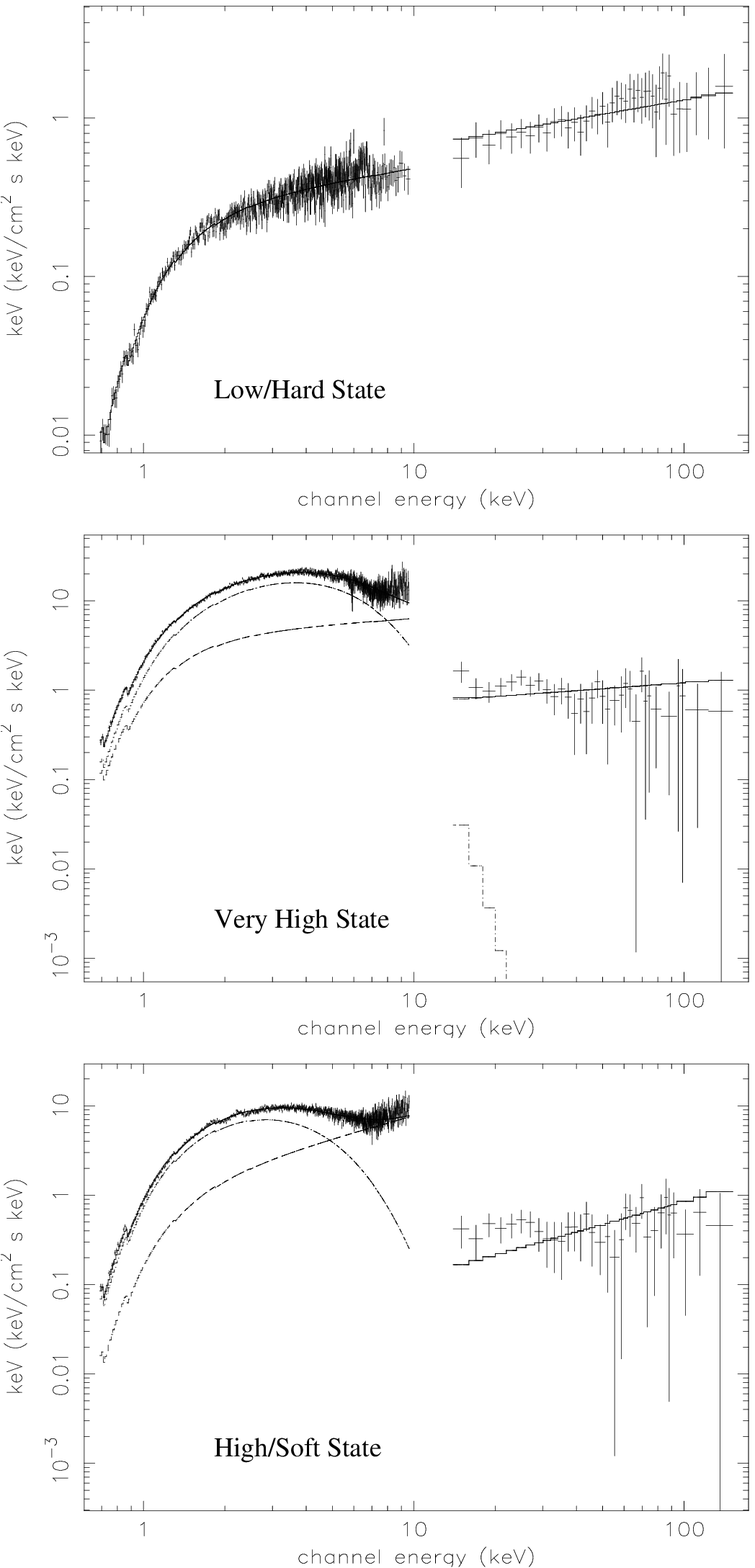,width=6cm}
\caption{Selected BAT--XRT spectra (ID 00030009002, 00030009020, 00030009025) showing the low/hard (top), very high (middle) and high/soft (bottom) states and plotted in E*E*f(E)-space (corresponding to $\nu F_{\nu}$). They have been fit with a power-law and, in the latter two, an additional disc blackbody (diskbb in {\sc xspec}). Fit parameters are listed in Table~\ref{tab:xraymodel}, although we note that the BAT spectra have been rebinned to a significance of $\ge 2\sigma$ in order to improve the plots.} 
\label{fig:xray-spectra}
\end{center}
\end{figure}


These models provide reasonably successful fits for the three observations plotted. The power-law model used to fit the initial low/hard state observation yields a photon index of $1.72\pm0.03$, consistent with quasi-simultaneous {\sl RXTE}/PCA observations (Homan 2005) and also consistent with the photon index expected for the low/hard state (McClintock \& Remillard 2005). We note that, on rebinning the spectrum to 300 counts and comparing the relative 90\% confidence limits to the line and continuum normalisations, it is also possible to include an iron line component of fixed width 0.01 keV in the fit at $6.39^{+0.05}_{-0.07}$ keV. This improves the $\chi ^2$ by $4.6\sigma$, compared with a power-law alone; the equivalent width of this line is $69^{+34}_{-36}$ eV (errors are to 90\% confidence). The photon indices of the subsequent spectra are much less well-defined; this is not surprising since the majority of the (now higher) flux is contained in the black-body disc component. There was also a problem with the XRT calibration at the high-energy end of the spectrum, resulting in negative photon indices for some observations; in these cases we lower the upper energy limit to 7.5-8.0 keV. However, the high/soft and very high state observations yield disc temperatures predominently in the range 1.3--1.5 keV, as expected for these two spectral states and in agreement with the {\sl RXTE} (Homan et al. 2005). The LrPD-mode fits are also affected by contamination by the calibration sources in the 5.5--7.0 keV range. However, preliminary detailed analysis of the background suggests that, during the high/soft and very high states, absorption features are present at $6.78\pm0.03$ and $7.07\pm0.03$ keV. We note that this is consistent with the results of Ueda et al. 1998. Furthermore, rebinning of the later WT spectra (during the second high/soft state) to 300 counts also reveals the presence of an absorption line, this time at the lower energy of 6.4--6.6 kev but still with significance of $\ge 4\sigma$. We save further line analysis for a future work.



\begin{table*}
\caption{Results of fitting simple spectral models to the combined BAT and XRT data. Errors and upper limits are to 90\% confidence. The columns are as for Table~\label{tab:xrtmodel} with the exception of Column 8, which gives the relative normalisation between the XRT and BAT. Again, due to calibration uncertainties at XRT high energies resulting in a negative power-law slope, it was necessary to restrict the XRT energy range for the observation marked *.}
\begin{tabular}{lcccccccccc}
\hline
\hline
Obs. ID &Model 	& nH 	&Photon &PL Norm. &Disc Temp. &BB Norm.&XRT/BAT &$\chi^2$&$\chi^2_{\nu}$&D.O.F\\
          &           &($\times10^{22}\mbox{cm}^{-2}$)&Index&&(KeV)&&Norm.&&&\\
\hline
00030009002 &PL     & $0.58^{+0.02}_{-0.01}$ & $1.70^{+0.03}_{-0.03}$ & $0.24^{+0.01}_{-0.01}$   & --                     & --                  & $1.34^{+0.14}_{-0.11}$ & 700.8  & 1.10 & 632\\
00030009005 &PL + BB& $0.69^{+0.01}_{-0.02}$ & *$1.70^{+0.11}_{-0.13}$&  $3.06^{+0.58}_{-0.64}$   & $1.36^{+0.02}_{-0.02}$ & $705^{+56}_{-47}$   & $0.06^{+0.01}_{-0.01}$ & 1275.0 & 1.62 & 786\\ 
00030009006 &PL + BB& $0.59^{+0.02}_{-0.07}$ & $1.71^{+0.59}_{-0.69}$ & $1.43^{+5.47}_{-1.40}$   & $1.51^{+0.14}_{-0.24}$ & $512^{+314}_{-200}$ & $0.06^{+0.27}_{-0.05}$ & 95.4   & 0.33 & 289\\
00030009007 &PL + BB& $0.69^{+0.03}_{-0.02}$ & $2.12^{+0.27}_{-0.18}$ & $4.27^{+0.86}_{-0.73}$   & $1.43^{+0.05}_{-0.04}$ & $491^{+50}_{-41}$   & $0.03^{+0.06}_{-0.03}$ & 711.2  & 0.80 & 890\\
00030009008 &PL + BB& $0.78^{+0.02}_{-0.03}$ & $3.29^{+0.20}_{-0.21}$ & $7.67^{+0.86}_{-0.77}$   & $1.56^{+0.01}_{-0.02}$ & $245^{+14}_{-15}$   & $0.54^{+0.56}_{-0.51}$ & 953.6  & 1.07 & 892\\
00030009011 &PL + BB& $0.68^{+0.03}_{-0.02}$ & $2.55^{+0.27}_{-0.18}$ & $6.28^{+1.80}_{-0.77}$   & $1.43^{+0.03}_{-0.04}$ & $400^{+31}_{-24}$   & $0.12^{+0.11}_{-0.07}$ & 808.3  & 0.98 & 828\\
00030009012 &PL + BB& $0.66^{+0.04}_{-0.03}$ & $2.17^{+0.25}_{-0.20}$ & $5.64^{+1.45}_{-1.37}$   & $1.39^{+0.07}_{-0.06}$ & $489^{+97}_{-75}$   & $<0.04$                & 811.1  & 1.11 & 728\\
00030009014 &PL + BB& $0.66^{+0.03}_{-0.02}$ & $2.31^{+0.35}_{-0.21}$ & $4.70^{+1.12}_{-0.80}$   & $1.49^{+0.04}_{-0.04}$ & $438^{+39}_{-27}$   & $0.06^{+0.07}_{-0.01}$ & 1099.0 & 1.25 & 878\\
00030009015 &PL + BB&  -- & --                         & --                              & --                             & --                  & --                     & 2259.0 & 2.39 & 945\\
00030009017 &PL + BB& $0.68^{+0.01}_{-0.01}$ & $1.42^{+0.05}_{-0.05}$ & $7.20^{+0.75}_{-0.64}$   & $1.23^{+0.02}_{-0.01}$ & $1154^{+87}_{-74}$  & $0.04^{+0.01}_{-0.01}$ & 1882.3 & 1.99 & 946\\          
00030009018 &PL + BB& $0.61^{+0.01}_{-0.01}$ & $1.42^{+0.08}_{-0.07}$ & $4.06^{+0.70}_{-0.55}$   & $1.27^{+0.02}_{-0.01}$ & $974^{+55}_{-69}$   & $0.04^{+0.01}_{-0.01}$ & 1814.8 & 1.92 & 946\\
00030009019 &PL + BB& $0.64^{+0.01}_{-0.01}$ & $1.78^{+0.05}_{-0.07}$ & $4.56^{+0.47}_{-0.47}$   & $1.40^{+0.02}_{-0.02}$ & $579^{+40}_{-36}$   & $0.10^{+0.02}_{-0.01}$ & 1369.3 & 1.45 & 945\\
00030009020 &PL + BB& $0.65^{+0.01}_{-0.02}$ & $1.78^{+0.13}_{-0.17}$ & $3.86^{+0.40}_{-0.97}$   & $1.42^{+0.04}_{-0.04}$ & $614^{+79}_{-55}$   & $0.11^{+0.04}_{-0.03}$ & 968.7 & 1.15 & 842\\
00030009021 &PL + BB&  --                    & --                     & --                       & --                     & --                  & --                     & 2183.2 & 2.35 & 928\\
00030009023 &PL + BB& $1.10^{+0.01}_{-0.02}$ & $1.83^{+0.02}_{-0.01}$ & $12.90^{+0.30}_{-0.30}$  & $<0.18$ & $<8.9$              & $0.06^{+0.01}_{-0.01}$ & 1518.3 & 1.68 & 906\\
00030009025 &PL + BB& $0.74^{+0.02}_{-0.01}$ & $1.15^{+0.21}_{-0.26}$ & $1.09^{+0.62}_{-0.45}$   & $1.01^{+0.02}_{-0.02}$ & $1175^{+63}_{-72}$  & $0.02^{+0.01}_{-0.01}$ & 1200.5 & 1.44 & 831\\

\hline 
\label{tab:xraymodel}
\end{tabular}
\end{table*}

\subsection{Broad-band spectra}

The power-law spectral components of the low/hard and very high states have been distinguished in terms of their relative steepness and high energy cut-off values. The hard state is expected to exhibit a photon index of $\sim $ 1.7, up to 100 keV; the very high state does not show such a cut-off, its significantly steeper spectrum ($\sim 2.5-3.0$) instead extending to MeV values, hence the alternative name ``steep power-law state'' (McClintock \& Remillard 2005). 

The power-law indices obtained by fitting the XRT and BAT spectra together are consistent with these values and are listed in Table~\ref{tab:xraymodel}. During the high/soft state the photon-indices appear softer (as we would expect), although the spectrum appears to harden during the very high state. However, we recommend some caution in interpreting these results since the parameters listed in Table~\ref{tab:xraymodel} are, in some cases, perhaps more representative of the lower signal/noise and the need for improved calibration than the true fit quality. Interestingly, the combined XRT--BAT photon indices show steeper spectra than the BAT alone, possibly indicating curvature; improved calibration, particularly at the high energy of the XRT range, would be needed to confirm this. Plots showing one spectrum for each spectral state are shown in Fig.~\ref{fig:xray-spectra}; we note that we have rebinned the BAT spectra to a significance of $\ge 2\sigma$ for each plot.

\begin{figure}
\begin{center}
\leavevmode
\epsfig{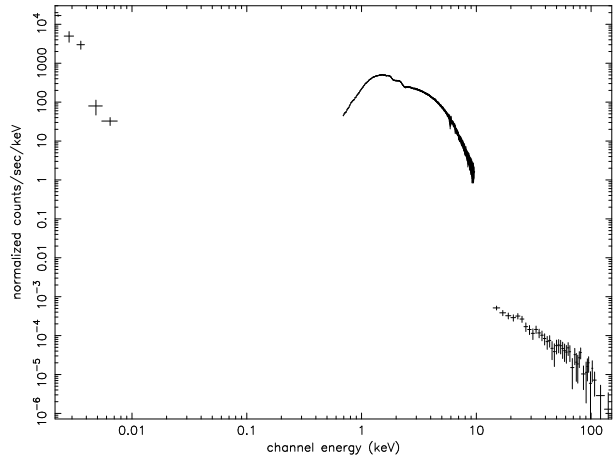}
\caption{Sample broad-band UVOT--XRT--BAT spectral energy distribution (ID 00030009019) included for completeness. Detailed fitting of the complete SED is beyond the scope of this paper but will be performed at a later date with improved calibration.}
\label{fig:sed}
\end{center}
\end{figure}

For completeness we also plot a sample UVOT--XRT--BAT spectrum in Fig.~\ref{fig:sed}. Future work will include detailed fitting of the broad-band spectral energy distribution in the more physical $\nu F_{\nu}$-space, once improved calibration is available.

\section{Discussion}

If we compare the lightcurve morphology of the 2005 outburst of GRO J1655$-$40 with outbursts of previous ``soft X-ray transients'' then we might be tempted to conclude that there was something strange about it; there is little evidence for the Fast Rise Exponential Decay (FRED) often considered canonical for ``soft X-ray transients'' (e.g. Lasota 2001 and references therein). However, not only have the majority of recent well-sampled lightcurves not shown the FRED morphology, the {\em spectral} evolution of the outbursts {\em has} been comparable with the current outburst of GRO J1655$-$40.

The recent {\sl Swift} and {\sl RXTE} observations have shown that GRO J1655$-$40 first rose from quiescence and entered the low/hard state. It then softened and entered the high/soft state for a few weeks, before becoming brighter and more variable and reaching the very high state. The very high state appeared to be associated with brief phases of temporary hardening.

This spectral behaviour has been observed in other objects and seems to be a much more predictable feature than the lightcurve morphology. XTE J1550$-$564 is an interesting source for comparison with GRO~J1655$-$40 since both have shown multiple outbursts, have been well-monitored throughout each of the spectral states and have been observed to eject jets with apparent superluminal motion (e.g. Homan et al. 2000; Sobczak et al. 2000; Corbel et al. 2001; Hannikainen et al. 2001 and references therein). The lightcurve morphologies, on the other hand, have been much less comparable; the spectral information is essential.

In particular the initial low/hard state was studied by Brocksopp et al. (2002a, 2004) and has been shown to be present in outbursts of all objects for which early data was obtained; evidence for the flat-spectrum, compact jet associated with the low/hard state was also found in each case that data were available. The duration of this state varies from source to source; a few days is typical, although XTE~J1650$-$500 remained hard for $\sim$ 1 month before softening and some sources have remained hard throughout their outbursts (Brocksopp et al. 2004). Regardless of when the softening takes place, the initial low/hard state occurs with such consistency that its role should be considered vital to our understanding of the outburst trigger and mechanism.

Interestingly the 1996 outburst of GRO~J1655$-$40 seemed to be a possible exception to the otherwise almost ubiquitous initial low/hard state. It was notable for its very soft rise, with an optical precursor and delayed hard X-rays relative to the soft X-rays (Hynes et al. 1998; Orosz et al. 1997). This was very different behaviour from either the first observed outburst in 1994, or the current activity. However, as there was only an interval of $\sim$ 8.5 months between the end of the 1995 activity and the 1996 rise, we do have to consider whether or not they were truly independent outbursts. The 1994 outburst was indeed hard at the onset, although the lack of soft X-ray monitoring makes it difficult to compare in detail.

The periods of temporary hardening have also been observed previously, although not as consistently as the initial low/hard state. XTE J1859+226 was a particularly good example since the temporary hardenings took place on a generally softening spectrum and simultaneously with radio ejections (Brocksopp et al. 2002a). Observing coverage has not generally been sufficient to see whether we should always {\em expect} radio ejections simultaneously with X-ray temporary hardenings during the very high state. We save this discussion for subsequent multi-observatory analysis.

The quasi-correlated behaviour between the hard and soft X-rays of the 2005 outburst and the lack of correlation between the optical/ultraviolet and the X-ray lightcurves can possibly be interpreted in terms of the two separate components which are generally used for interpretation of the spectra:

\begin{enumerate}
\item An accretion disc component which brightens smoothly during the rise of the outburst and then drops once the source reaches the very high state, providing the optical/ultraviolet lightcurve and the gradually-rising part of the soft X-ray lightcurve. This is presumably related to the disc black-body component, typically used when fitting the X-ray spectrum.
\item A jet and/or corona component which becomes active at the onset of the outburst, peaks and decays and then rebrightens suddenly, resulting in the very high state. Again, this is presumably related to the power-law component (either synchrotron or Comptonisation model), typically used when fitting the X-ray spectrum.
\end{enumerate}

Such a model could then explain the apparent lack of correlation between the hard X-ray, soft X-ray and optical/UV frequency regimes, although the often-cited ``disk/jet/corona'' connection is still vital in providing the link between the two components (for example producing the drop in optical/ultraviolet flux during the very high state). The apparent absence of the expected FRED-shaped lightcurve could also be explained -- it {\em is} present but at times is dominated by the non-thermal, non-FRED component. The lightcurves of the 1996 outburst suggest a similar scenario, although with the exception that (as already discussed) the accretion disc component became active before the jet/corona component. This could explain why the 1996 soft X-ray lightcurve showed a rise and only {\em slight} decay while the optical emission rose and decayed, but then rebrightened once the hard X-rays and radio source became active. In further support of this speculative model we note that an increase in flux via a thermal mechanism would be expected to take place at progressively higher frequencies (an ``outside--in'' outburst), as observed by Orosz et al. (1997). However, jet models predict the opposite behaviour (e.g. Blandford \& K\"onigl 1979; Brocksopp et al. 2002b), consistent with the delayed soft X-ray emission observed in both the 1996 and 2005 outburst of GRO J1655$-$40. We also note that Rupen et al. (2005c) report a radio peak on MJD 53494.5, {\em prior} to the X-ray peak; clearly the situation is more complicated, requiring detailed radio/X-ray analysis.

\section{Conclusions}
{\sl Swift} observed the 2005 outburst of GRO J1655$-$40, using all three of its instruments, and monitored the evolution of the source through the low/hard, the high/soft and very high spectral states. The spectra were modelled with an absorbed power-law and (for the two softer states) an additional multicolour disc black-body component; the parameters were typical for these ``canonical'' states. The inclusion of an iron line in the low/hard state spectrum improved the fit by $4.6\sigma$, although whether the line is a feature of the low/hard state alone or just dominated by the black-body in the softer states is currently unclear. The lightcurves showed that the high energy X-ray flux peaked before that of the soft X-rays and that there was no apparent correlation between the X-ray and optical/ultraviolet behaviour. We suggest that the soft X-ray lightcurve is a composite of emission contributions from a thermal component, correlated with the optical/ultraviolet, and from a non-thermal component, correlated with the hard X-rays. Such a model may also be consistent with the soft X-ray delays observed during both the 2005 and the 1996 outbursts.

\section*{acknowledgments}
This work is sponsored at PSU by NASA's Office of Space Science through grant NAG5-8401, and at MSSL by funding from PPARC. We thank Dave Morris for useful advice regarding the calibration for the XRT. CB, KEM and OG gratefully acknowledge funding through PPARC.


\begin{thebibliography}{}
\bibitem[]{}Abbey A., Stevenson T., Ambrosi R., 2005, Proceedings of ``The X-ray Universe'' Conference, El Escorial Spain, 2005, in prep.  
\bibitem[]{}Arnaud K.A et al., 1996, Astronomical Data Analysis Software and Systems V, A.S.P. Conference Series, Vol. 101, 1996, George H. Jacoby and Jeannette Barnes, eds., p. 17
\bibitem[]{}Bailyn C.D. et al., 1995, Nature, 374, 701
\bibitem[]{}Barthelmy S. et al., 2005, Space Science Reviews in press (astroph/0507410)
\bibitem[]{}Beer M.E., \& Podsiadlowski P., 2002, MNRAS, 331, 351
\bibitem[]{}Blandford R.D., K\"onigl A., 1979, ApJ, 232, 34
\bibitem[]{}Brocksopp C. et al., 2002a, MNRAS, 331, 765 
\bibitem[]{}Brocksopp C., Fender R.P., Pooley G.G., 2002b, MNRAS, 336, 699
\bibitem[]{}Brocksopp C., Bandyopadhyay R.M., Fender R.P., 2004, NewA, 9, 249
\bibitem[]{}Burrows D.N. et al., 2005, Space Science Reviews in press (astroph/0508071)
\bibitem[]{}Buxton M., Bailyn C., 2005, ATel 485
\bibitem[]{}Corbel S. et al., 2001, ApJ, 554, 43
\bibitem[]{}Fender R.P., 2005, Chapter 9 in ``Compact Stellar X-ray Sources'', Eds. W. Lewin \& M. van der Klis, CUP
\bibitem[]{}Fender R.P., Belloni T.M., 2004, ARA\&A, 42, 317
\bibitem[]{}Gehrels N. et al., 2004, ApJ, 611, 1005
\bibitem[]{}Hannikainen D.C., Hunstead R.W., Campbell-Wilson D., Wu K., McKay D.J., Smits D.P., Sault R.J., 2000, ApJ, 540, 521
\bibitem[]{}Hannikainen D.C. et al., 2001, ApJSS, 276, 45
\bibitem[]{}Harmon B.A. et al., 1995, Nature, 374, 703
\bibitem[]{}Hill J.E., 
\bibitem[]{}Hjellming R.M., Rupen M.P., 1995, 375, 464
\bibitem[]{}Homan J., 2005, ATel 440
\bibitem[]{}Homan J., Belloni T.M., 2005, in From X-ray Binaries to Quasars: Black Hole Accretion on All Mass Scales, ed. T. J. Maccarone, R. P. Fender, L. C. Ho (Dordrecht: Kluwer). (astro-ph/0412597)
\bibitem[]{}Homan J., Miller J.M., Wijnands R., Lewin W.H.G., 2005, ATel 487
\bibitem[]{}Homan J., Wijnands R., van der Klis M., Belloni T., van Paradijs J., Klein-Wolt M., Fender R., M\'endez M., 2000, ApJS, 132, 377
\bibitem[]{}Hynes R.I., 1998, MNRAS, 300, 64
\bibitem[]{}Lasota J.-P., 2001, NewAR, 45, 449  
\bibitem[]{}Markwardt C.B., Swank J.H., 2005, ATel 414
\bibitem[]{}Markwardt C.B., Homan J., Swank J.H., 2005, ATel 436
\bibitem[]{}McClintock J., Remillard R., 2005, Chapter 4 in ``Compact Stellar X-ray Sources'', Eds. W. Lewin \& M. van der Klis, CUP
\bibitem[]{}Mirabel I.F., Rodr\'{\i}guez L.F., 1994, Nature, 371, 46
\bibitem[]{}Moretti A., et al., 2005 SPIE Proc., Vol 5898, in press
\bibitem[]{}Orosz J.A., Bailyn C.D., 1997, ApJ, 477, 876
\bibitem[]{}Orosz J.A., Remillard R., Bailyn C.D., McClintock J.E., 1997, ApJ, 478, L83
\bibitem[]{}Roming P.W.A. et al., 2005, Space Science Reviews in press (astroph/0507413)
\bibitem[]{}Rupen M.P., Dhawan V., Mioduszewski A.J., 2005a, ATel 419
\bibitem[]{}Rupen M.P., Mioduszewski A.J., Dhawan V., 2005b, ATel 434
\bibitem[]{}Rupen M.P., Dhawan V., Mioduszewski A.J., 2005c, ATel 489
\bibitem[]{}Smith D.A., 2005, ATel 486
\bibitem[]{}Sobczak G.J., McClintock J.E., Remillard R.A., Cui W., Levine A.M., Morgan E.M., Orosz J.A., Bailyn C.D., 2000, ApJ, 544, 993
\bibitem[]{}Tavani M., Fruchter A., Zhang S.N., Harmon B.A., Hjellming R.M., Rupen M.P., Bailyn C., Livio M., 1996, ApJ, 473, L103
\bibitem[]{}Tingay S.J. et al., 1995, Nature, 374, 141
\bibitem[]{}Ueda Y., Inoue H., Tanaka Y., Ebisawa K., Nagase F., Kotani T., Gehrels N., 1998, ApJ, 492, 782
\bibitem[]{}Van der Klis M., 2005, Chapter 2 in ``Compact Stellar X-ray Sources'', Eds. W. Lewin \& M. van der Klis, CUP
\bibitem[]{}Zhang S.N. et al., 1997, ApJ, 479, 381

\end{thebibliography}
\end{document}